# Microfluidic Lab-on-a-Chip characterization of nano- to microparticles suspensions by light extinction spectrometry

**FABRICE R.A. ONOFRI,[1] ISAAC RODRIGUEZ-RUIZ,[2] AND FABRICE LAMADIE,[3]**

[1] *Aix-Marseille University, CNRS, IUSTI, UMR 7343, Marseille, France*
[2] *Laboratoire de Génie Chimique - CNRS, UMR 5503, 4 Allée Emile Monso, Toulouse, France*
[3] *CEA, DES, ISEC, DMRC, Univ Montpellier, 30207 Bagnols-sur-Ceze, Marcoule, France*
[1]fabrice.onofri@univ-amu.fr
[2]isaac.rodriguezruiz@toulouse-inp.fr
[3]fabrice.lamadie@cea.fr

**Abstract:** The analysis of nano- and microparticle suspensions with micro systems affords improved space–time yields, selectivity, reaction residence times and conversions capabilities. These capabilities are of primary importance in various fields of research and industry. The few microfluidic lab-on-a-chip approaches that have been developed are essentially designed to analyse fluid phases or involve the use of benchtop particle sizing instruments. We report a novel microscale approach to characterize the particle size distribution and absolute concentration of colloidal suspensions. The method is based on a photonic lab-on-a-chip with three scale-specific detection channels to record simultaneous light extinction spectra. Experiments carried out on particle standards with sizes ranging from 30 nm to 0.5 µm and volume concentrations of 1 to 1000ppm, clearly demonstrate the value and potential of the proposed method.

## 1. Introduction

Microfluidics has become an essential tool in many areas from life sciences [1, 2] to chemistry [3] and environmental analysis [4]. In chemical engineering [5], microfluidics improves the space–time yields, selectivity, reaction residence times and conversion performances of fundamental synthetic transformations in solution. This makes possible, for instance, to synthetize cadmium sulphide [6, 7], gold [8], titania [9], or CdSe–ZnS [10] nanoparticles of controlled sizes and morphologies. Nevertheless, several challenges remain to be overcome before microfluidics becomes a truly multipurpose and practical tool. One of the main current difficulties is obtaining real-time, high-performance, quantitative information on processes at the same micrometre scale [11]. Most lab-on-a-chip (LoC) setups still rely on bulky and expensive benchtop instruments for analyte/product characterization. This limits their attractiveness compared with standard protocols and techniques, notably in real-world applications outside controlled laboratory environments [12]. In the quest for compact analytical platforms, different detection and characterization systems have already been integrated in LoC devices, based either on electrochemical [13], magnetic [14], or optical [15] transducers. While these devices have already been successful in the characterization of continuous (fluid) phases or effective media [12, 16-18], their development for the characterization of disperse phases (droplets, bubbles, crystals or aggregates) is still in its infancy. Classical approaches based on the temporal coherence of light (Dynamic Light Scattering, DLS) are limited in terms of flow dynamics and temperature control, while those based on angular scattering (Multi-Angle Static Light Scattering, MASLS) are difficult to implement at the microscale and limited to microparticles. Additionally, it is worth noting that both approaches do not provide absolute particle concentrations [19-23]. To address these issues, we report a novel approach to characterize particle size distributions (PSDs) and concentrations of nano- to microparticles in static as well as flowing suspensions. This approach is based on the principle of light extinction spectrometry (LES) [24-36] and a photonic lab-on-a-chip with three scale-probing channels, see Figure 1. After this brief introduction, section 2 summarizes the background and basic requirement of LES. Section 3 presents the testing procedure and experiment set up to evaluate this first prototype on nano- and microparticle calibration standards, while the results are presented and discussed in section 4. Section 5 is a general conclusion with perspectives.

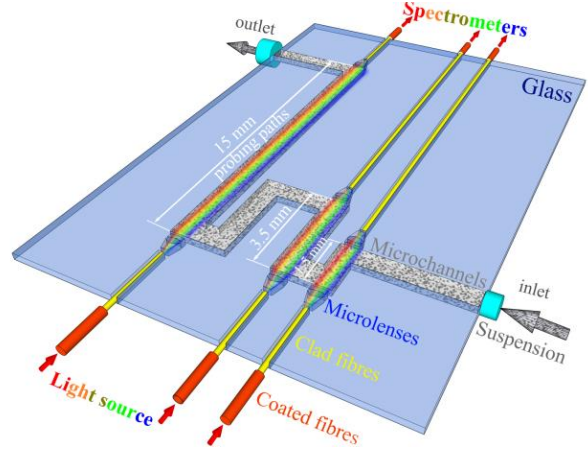

Figure 1 Schematic diagram of the photonic LoC, with three channel paths and light extinction capabilities over a broadband spectral range, for characterizing the particle size distribution and absolute concentration of colloidal suspensions.

## 2. LES background

LES basically involves recording the spectral transmittance $T_\lambda = I_\lambda^P / I_\lambda^0$ of a broadband collimated light beam with incident and final intensities $I_\lambda^0$ and $I_\lambda^P$, respectively, after its passage through the analysed particulate medium. These spectra are compared via a minimization process to those predicted by light absorption and scattering models of particle extinction (based for example on Lorenz-Mie theory (LMT) for spherical and multilayered particles [37] or the T-Matrix method for spheroids and aggregates [31, 38]). The basic LES equation for a homogeneous and optically dilute particulate medium with particle concentration, $C_v$, a dimensionless quantity, and volume-normalized PSD, $v_D$ is [32-34]:

$$\ln\left(T_\lambda^{th}\right) = -\frac{3LC_v}{2} \int_{D_{\min}}^{D_{\max}} \left(Q_{\lambda, m_\lambda, D} / D\right) v_D dD, \qquad (1)$$

where $Q_{\lambda, m_\lambda, D}$ is the extinction efficiency factor of a particle with diameter D and relative refractive index $m_\lambda = m_{\lambda,0}/m_{\lambda,s}$; $m_{\lambda,0}/m_{\lambda,s}$ is the ratio of the refractive indexes of the particle and the solvent at the wavelength in air, $\lambda_o$; $\lambda = \lambda_o/m_{\lambda,s}$ is the local wavelength; and L is the optical path length. The discretized and algebraic forms of this equation, its variables and the reconstructed transmission are detailed in [32]. Eq. 1 is indeed the linearized and generalized form of the Beer-Lambert equation which, for a monodisperse sample of spherical particles, is simply:

$$T_\lambda^{th} = \exp\left[-3LC_v Q_{\lambda, m_\lambda, D} / (2D)\right]. \qquad (2)$$

Eq. 1 is also a Fredholm equation of the first kind, such that LES is an inverse method [28, 33]. Controlling the optical path length is crucial because as shown by Eq. (2), the dynamic range and sensitivity of LES are exponentially dependent on this parameter. The harmful effects of multiple scattering can also be detected, to some extend, by comparing the transmissions obtained with different path lengths. The difference between the measured and theoretical transmittances ($T_\lambda^{exp}$ and $T_\lambda^{th}$) can be minimized by least squares, $\left|T_\lambda^{th} - T_\lambda^{exp}\right|^2$ to determine the PSD and, afterwards, the particle concentration by volume (or number). If the minimization is performed by fixing certain parameters, this typically only yields the moments of the PSD (e.g., the standard deviation and mean value for a log-normal distribution) and possibly its boundaries. In contrast, full regularization methods (e.g., Phillips-Twomey or Tikhonov) allow the real shape of the distribution to be retrieved provided the data are not too noisy [28, 31, 33, 39].

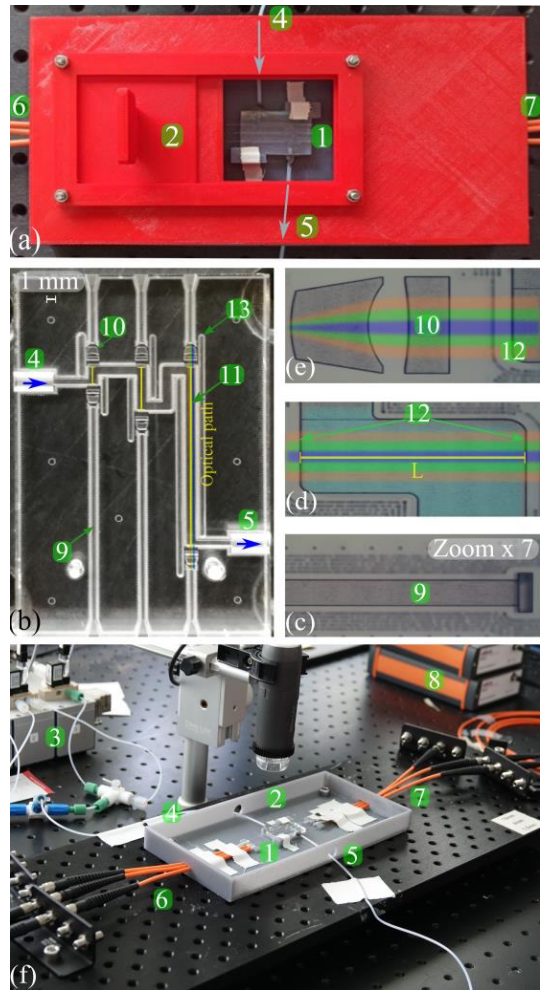

Figure 2 (a-f) Lab-on-a-chip: (1) Glass LoC in its (2) housing (hatch open); (3) computer controlled syringe pumps (x2) feeding the LoC via its (4) inlet and (5) outlet; suspension (4) inlet and (5) outlet; (6) incoming and (7) outgoing optical fibres; (8) the CCD spectrometers (x3); (c-e) Expanded views of (9) the self-alignment channels (x6) for optical fibres; (10) the 2D microlense assemblies (x6) for collimating and controlling optical fibre outputs and inputs; (11) the optical channels (x3) with path lengths L=1.5, 3.5 and 15 mm, width 650 µm and depth 250 µm; (12) the ends of the channels with parallel optical-grade windows; and (13) air channels to avoid probing channels cross-talking during simultaneous acquisition. The scratches visible in part (b) are those on the breadboard underneath the LoC. False colours were added in parts (b, d, e) to highlight the paths of the fluids (light blue) and probing beams (multicoloured). (f) Setup with (2) LoC housing cover removed.

Another key assumption of LES is that the amount of scattered light detected (scattered by the sample or any component of the optical setup) is negligible [24, 32]. In practice, this sets upper limits on the size of the particles and the optical density of the medium. Inelastic scattering processes [40] and polarization effects [41, 42] are also neglected in the standard working principles of LES. Knowledge of the composition and morphology of the particles are also required, even if both can be inferred in some situations [26, 31, 33, 35, 36]. Although LES is mainly used to characterize particulate media in large facilities (e.g. nanocrystallite nucleation and growth in plasma [32, 34], fractal aggregation in aerosols or deposited on an interface [30, 33]), it has two considerable advantages for the microscale characterization of nano- to microsuspensions. First, its detection scheme is rather simple and fast, and its sensitivity can be controlled via the optical path length. Second, and crucially for many processes and studies, LES provides absolute measurements of particle concentrations, in stationary as well as flowing or transient suspensions.

### 3. Experimental setup and procedure

We designed a glass LoC with three sensing microchannels of different optical path lengths (L = 1.5, 3.5 and 15 mm respectively), with width 650 µm and depth 250 µm, see Figure 1 and Figure 2). Glass was chosen, over PDMS for instance [12], because of its optical and mechanical properties regarding, notably, chemical exposure and irradiation [43]. This monolithically integrated device was manufactured by FEMTOPRINT® following our specifications, with an innovative femtosecond laser pulses micromachining technology for ultra-fast micromolding of fused silica. With a high repeatability and alignment precision (< 1µm) and a low arithmetical mean deviation of the assessed profile (< 100nm), a surface accuracy of better than $\lambda/10$ is obtained after additional chemical etching. The glass LoC includes six self-aligned channels for optical fibres and six self-aligned pairs of 2D microlenses (with an effective focal length of 925 µm, optimized with a ray tracing procedure introduced in a previous work [18]) for beam collimation and detection angle control. The channels are all 650 µm wide and 250 µm deep to limit the amount of suspension required for the analyses. The optical channels end with optical grade parallel windows. Light is transmitted to the LoC and the spectrometers along solarization-resistant fibres (200 µm core) with a low numerical aperture (NA=0.22).

The three channels are lighted simultaneously by splitting the output of a fibre-optic balanced deuterium-halogen light source (A AvaLight-DH-S-BAL from Avantes, not shown for the sake of clarity). The spectra are recorded simultaneously with three integrated Czerny-Turner spectrometers equipped with high resolution and high sensitivity linear CCDs (AvaSpec-ULS4096CL-EVO from Avantes). As is commonly done, the signal-to-noise ratio (SNR) of the measured transmittance was improved by reducing the CCD exposure time as much as possible (to 11–17 ms depending on the optical path) and averaging 10 successive spectra, giving a measurement rate of nearly 6 per second. For the sake of simplicity, the potential of LES for LoC microfluidics was evaluated using six monodisperse particle standards provided by Thermo Scientific with NIST certificates based on SEM analyses of dry samples. The six standards are aqueous suspensions of polystyrene nano- and microparticles with mean diameters of 30 nm, 50 nm, 80 nm, 0.20 µm and 0.40 µm, and silica ($SiO_2$) microparticles with mean diameter 0.50 µm, with nominal concentrations in mass ranging from 0.1 to 1.0 wt%. Polystyrene suspensions with a maximum concentration range of 1, 5, 10, 50, 100, 150, 250, 400, 500, 750 and 1000 ppm were then prepared by successive dilutions with Type I ultrapure water obtained by distillation, deionization and reverse osmosis (Fisher Scientific Accu100). Concentrations of silica suspensions were limited to 50, 100, 250 and 500 ppm. Because of the high concentration of the initial suspensions and the successive dilutions, the estimated relative error on the concentration of the suspensions ranged from 3.5% for the more concentrated suspensions (above 50 ppm) up to 23 % for the 1 ppm suspensions. According to the manufacturer of the particle standards, the standard deviation at 2-σ of the PSDs is less than 3 nm for the polystyrene and 11 nm for the SiO$_2$ beads. The monodispersity of the samples was confirmed by benchtop DLS (Litesizer from Anton Paar) analyses (data not shown). Typically, each spectrophotometric acquisition consisted in recording spectra of the background ($I_\lambda^B$, light off), of the reference obtained with the solvent ($I_\lambda^S$, here water, light on) and of the particle sample ($I_\lambda^P$, light on), such that the experimental transmittance was calculated as

$$T_\lambda^{\exp} = \left(I_\lambda^P - I_\lambda^B\right) / \left(I_\lambda^S - I_\lambda^B\right). \tag{3}$$

To this end, the LoC was filled alternately with two computer-controlled syringes, one for the solvent and on for the analysed particle sample, after sonication. This procedure was repeated to obtain several transmittance signals for each suspension and channel. Spectral transmittances were analysed in a reduced spectral range (300–600 nm, out of a full spectrometer range of 195–1110 nm). Below 300nm, it was to avoid the effects of the unspecified mixture of surfactants and/or bases used as stabilizing additives and, above 600 nm, to prevent the detection of unexpected residual contributions from higher orders of the diffraction gratings. To account for the limited spectral resolution of the spectrometers, the theoretical transmittances $T_\lambda^{th}$ were convoluted (symbol $\circ$) as

$$T_\lambda^{th} \rightarrow T_\lambda^{th} \circ R_\lambda, \tag{4}$$

with the spectral-impulse-response $R_\lambda$ of the spectrometers, before performing the minimization process. This spectral-impulse-response $R_\lambda$ was estimated by analysing several laser outputs, take the form of Gaussian peaks whose standard deviations increase almost linearly from 3.6 to 4.1nm between 405 and 633 nm (extrapolated down to 300nm).

The data were analysed using LMT to calculate the kernel of Eq. 1, followed by least squares minimization assuming a normal PSD defined by its mean and standard deviation (i.e., parameters of the two-parameter model, as detailed in previous articles [32, 33]). The boundaries of the PSD, $D_{min}$ and $D_{max}$, were imposed at 1% of the mode [33]. For the refractive index dispersion of the materials (polystyrene, silica and water), we used relevant databases and softwares [44-46]. The minimization process was performed iteratively for mean sizes and standard deviations varied from 1 to 1000 nm and 1 to 50 nm (i.e., ranges of the two-parameter model), respectively, in 1 nm steps (for more details, see Ref. [33]). Several tests were used to reject transmittance signals with features that were (i) too weak ($T_\lambda^{\exp} < 0.05$, with a high risk of collecting scattered light) or (ii) too strong ($T_\lambda^{\exp} > 0.995$, potentially too noisy), or (iii) with an excessive high-frequency noise level (for previous reasons plus some effects of contaminants), or (iv) giving a reconstructed transmittance with a too high residual deviation (>3%) [33]. The validation rate of the testing procedure ranged from 0 to 80%, depending on the channel, the size and concentration considered, with a mean validation rate for all samples of 20%.

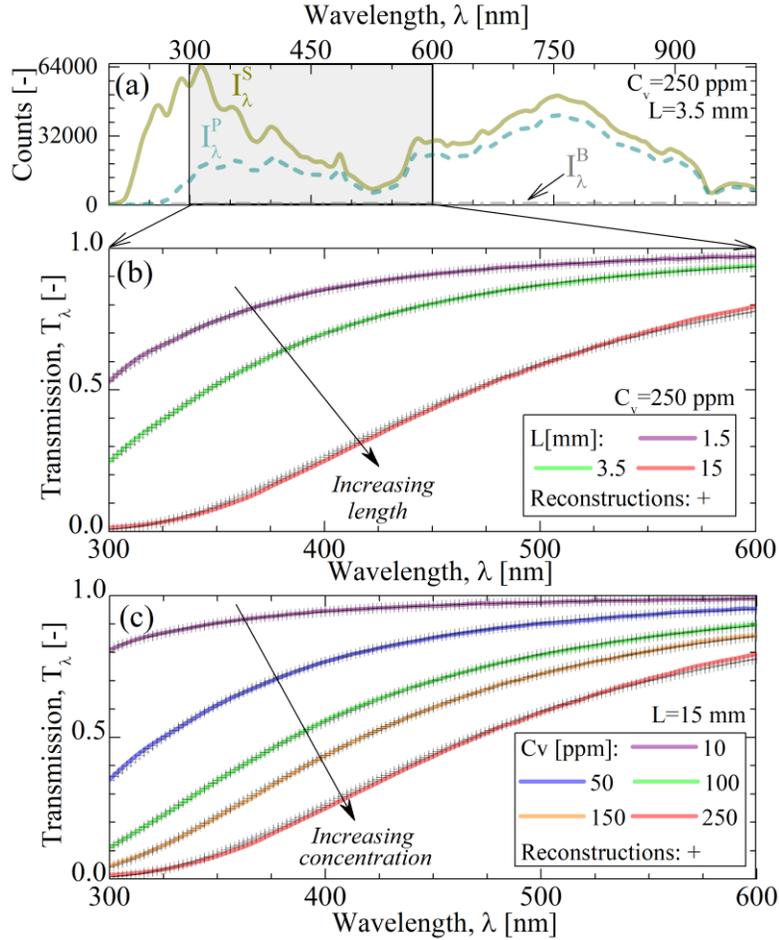

Figure 3 LES (a) raw signals; (b, c) experimental and reconstructed transmissions for suspensions of polystyrene nanoparticles of 80nm in diameters. (a) Typical raw signals for a concentration of 250 ppm and the 3.5 mm probing length. (b) Evolution of the transmissions for increasing channel path lengths when the particle concentration is fixed to 250 ppm. (c) Evolution of the transmissions for increasing particle concentrations when the optical path length is fixed to 15mm. The parameters estimation is performed on the spectral range 300–600 nm (highlighted in grey in (a)).

## 4. Results and discussion

Typical raw signals (reference, sample signal and background) are shown in Figure 3 (a), while Figure 3 (b) and Figure 3 (c) compare the experimental transmissions with those reconstructed for increasing channel path lengths and particle concentrations respectively. Figure 4 allows estimating the global resolution and dynamic range of all size measurements obtained with the device. The standard deviations of individual PSDs are not shown because they were below 1–4 nm at 1-$\sigma$. Each statistic is obtained from 10 to 55 measurements depending on the validation rate. Suspensions with particle sizes of 30, 50 and 400nm were measured on static samples, and the others, on flowing samples (50 µL/ min).

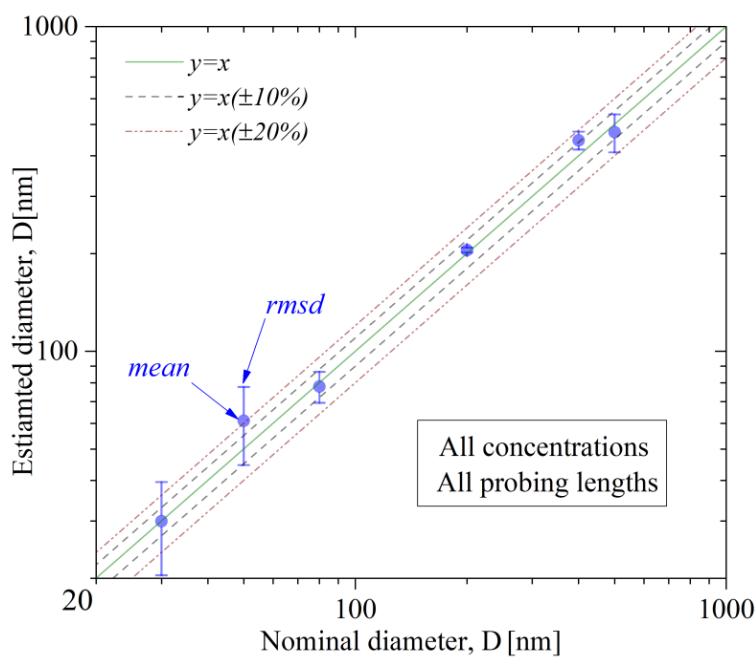

Figure 4 Mean diameter measured with all channels and for all prepared suspensions versus the mean diameter estimated by SEM analyses on the source suspensions. The error bars show the spread (rmsd) of the measured mean diameters. Static samples 30, 50 and 400nm, flowing samples: 80 and 500nm.

Looking at Figure 4, we found that the diameters measured with the photonic LoC are generally in good agreement with the reference values. The spread of the data, reflected by the length of the error bars (rmsd on all measured diameters for a given nominal diameter), is larger for the smallest particles. The fact is that smaller particles have lower extinction coefficients and thus higher and noisier transmittance. This is highlighted in Figure 3 which shows the transmittance signals for the 80 nm polystyrene particles at 5 and 250 ppm measured in the three channels. A part of the dispersion of the data may also be due to adventitious contamination of the samples, which were prepared and analysed in a conventional laboratory rather than in a clean room facility. Similarly, the water used for the dilutions was probably not completely free of nanocontaminants. The increase in the rmsd of the data for the largest diameters is harder to explain, except as being due to the detection of scattered light. Figure 5 compares the volume concentrations measured for all samples (validated signals only) with the expected values (i.e. deduced from manufacturer information and the dilution process). The measured concentrations are consistent with the expected values, despite a certain amount of dispersion that can be attributed to the same processes pointed out for the results in Figure 4. Each channel covers about two orders of magnitude in terms of suspension concentration. Such that in combination, the three channels can be used to analyse suspensions with concentration changes of up to three orders of magnitude, which is already promising for numerous applications. However, this dynamic range is not obtained for all particle sizes, depending on the SNR of the signals which highly depends on the optical path length but also on the quality of the suspension considered.

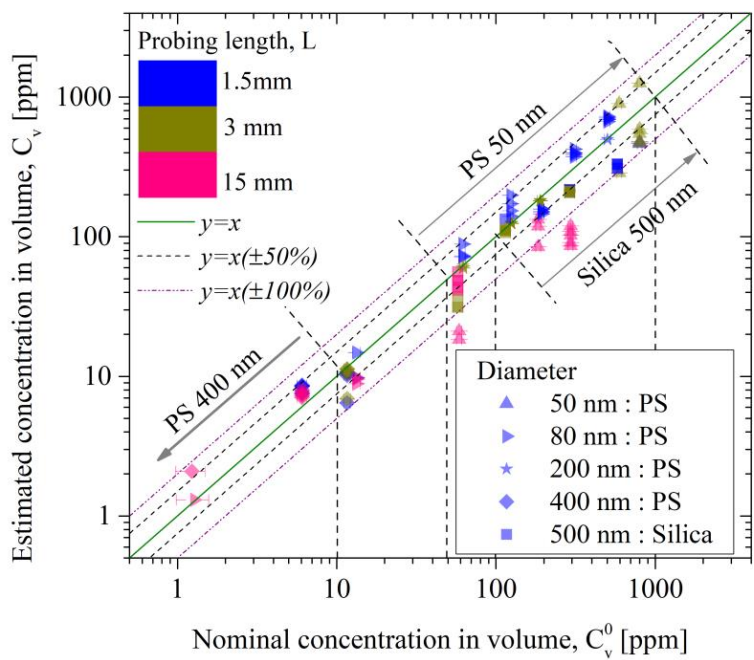

Figure 5 Volume concentration of aqueous suspensions of polystyrene (PS) and Silica particles versus the concentration calculated from the dilution process and the manufacturer-specified concentration of the starting suspension. Static samples: 30, 50 and 400nm, flowing samples: 80 and 500nm.

## 5. Conclusion and perspectives

We introduce and evaluate a new LoC approach to measure particle sizes (from 30 nm to 0.5 µm) and volume concentrations (from 1 to 1000 ppm) of optically dilute colloidal suspensions. This simple and effective spectrophotometric approach is applicable to both stationary suspensions and dynamic particulate flows, with potential applications in various fields of research and engineering. The design of the LoC, and more especially the detection channels, could be further optimized to fit the optical properties, size and concentration of the particles (i.e. the optical thickness of the samples) and to better control the SNR of the transmittance signals. Increasing substantially the optical path length [47] and integrating a microfluidic-controlled optical router [48] for multiplexed analyses with a single spectrometer would also be a benefit for practical applications. The consideration of particles with more complex shapes and compositions [30, 33, 35, 36], as well as multiple scattering or polarization effects [41, 42], is also part of the perspectives of the present work.

**Fundings.**

This research was supported by the Energy Division of the CEA (DES-DPE-CYN/PRATA) and was partially funded by the French National Research Agency (ANR) under grant numbers ANR-13-BS09-0008-02, Labex MEC (ANR-11-LABX-0092), and A*MIDEX (ANR-11-IDEX-0001-0).

**Disclosures.**

The authors declare no conflicts of interest.

**Data availability.**

The LES analysis software and the data that support the findings of this study are available from the authors upon reasonable request.